\documentclass[epjCONF]{svjour}
\usepackage{graphics}
\usepackage[varg]{txfonts} 
\usepackage[latin1]{inputenc}
\newcommand{\diamonds}{\textsc{D\large{iamonds}}}
\newcommand{\Kepler}{\textit Kepler}
\newcommand{\kepler}{{\textit Kepler}\,\,}
\newcommand{\numax}{\nu_\mathrm{max}}
\newcommand{\muhz}{$\mu$Hz}

\newcommand{\apj}{ApJ}

\newcommand{\aap}{A\&A}
\newcommand{\nat}{Nature}
\newcommand{\apjl}{ApJL}
\newcommand{\solphys}{Sol. Phys.}
\newcommand{\mnras}{MNRAS}
\newcommand{\kic}{KIC~12008916}
\newcommand{\evid}{\mathcal{E}}

\session-title{Conference Title, to be filled}

\begin{document}
\title{Peak Bagging of red giant stars observed by {\it Kepler}: first results with a new method based on Bayesian nested sampling}
\author{Enrico Corsaro\inst{1}\fnmsep\thanks{\email{emncorsaro@gmail.com}} \and Joris De Ridder\inst{1}}
\institute{Instituut voor Sterrenkunde, KU Leuven, Celestijnenlaan 200D, B-3001 Leuven, Belgium}
\abstract{
The peak bagging analysis, namely the fitting and identification of single oscillation modes in stars' power spectra, coupled to the very high-quality light curves of red giant stars observed by {\it Kepler}, can play a crucial role for studying stellar oscillations of different flavor with an unprecedented level of detail. A thorough study of stellar oscillations would thus allow for deeper testing of stellar structure models and new insights in stellar evolution theory.
However, peak bagging inferences are in general very challenging problems due to the large number of observed oscillation modes, hence of free parameters that can be involved in the fitting models. Efficiency and robustness in performing the analysis is what may be needed to proceed further. For this purpose, we developed a new code implementing the Nested Sampling Monte Carlo (NSMC) algorithm, a powerful statistical method well suited for Bayesian analyses of complex problems.
In this talk we show the peak bagging of a sample of high signal-to-noise red giant stars by exploiting recent Kepler datasets and a new criterion for the detection of an oscillation mode based on the computation of the Bayesian evidence. Preliminary results for frequencies and lifetimes for single oscillation modes, together with acoustic glitches, are therefore presented.
} 
\maketitle
\section{Introduction}
\label{intro}
Since the first detection of non-radial oscillations in cool giant stars \cite{DeRidder09}, the asteroseismology of field red giants (RGs) has encountered a substantial increase, especially thanks to the advent of the space-based photometric missions \textit{CoRoT} (e.g. \cite{Mosser11universal,Mosser11mixed}) and \kepler (e.g. \cite{Kallinger10Kepler,Kallinger12}), where the latter allowed also for the study of RGs in open clusters (e.g. \cite{Stello11membership,Corsaro12}).

The discovery of so-called mixed modes \cite{Beck11Science} has led to a significant improvement in our understanding of the internal structure and evolution of RGs. In particular, the characteristic spacing in period of mixed modes frequencies provides a direct way to disentangle H-shell and He-core burning RGs \cite{Bedding11Nature,Mosser12} and has been used already to classify about 13,000 targets observed in the \kepler field of view \cite{Stello13}. Therefore, RGs are among the most interesting and useful types of stars to refine models and test thoroughly codes of stellar structure and evolution.

With the present 4-years long datasets available from \kepler for many RGs, we have the possibility to perform very detailed analyses aimed at measuring asteroseismic properties such as frequency, amplitude, and lifetime, of each individual oscillation mode observed in the Fourier spectra of the stars. This particular analysis, usually known as peak bagging plays a key role to exploit the full potential of the informative power spectra of the stars. However, due to the large number of modes populating the power spectrum of a RG (up to $\sim$100), the peak bagging usually turns into a computationally demanding analysis \cite{Gruberbauer09,Kallinger10CoRoT,Handberg11,Corsaro14}. The complexity of the analysis is even higher considering that the mode identification of all the oscillation peaks is necessary for a proper modeling of the oscillations. This implies that also testing the significance of individual oscillation peaks becomes an important task within the process. For this purpose, \cite{Corsaro14} developed a new code termed \diamonds\,\,that implements the nested sampling Monte Carlo algorithm \cite{Skilling04}, a sampling algorithm well suited for model comparison and efficient high-dimensional inferences in a robust Bayesian perspective \cite{Corsaro14}.

\section{Observations and Data}
\label{sec:1}
To select the final sample of stars for the analysis we started from the original sample of RGs observed in long cadence (LC) by \kepler and studied by \cite{Huber11} and by \cite{Corsaro13}, consisting in a total of 1111 stars. 

From the original sample we only considered those stars with power spectra having an oscillation envelope with frequency of maximum power $\numax \gtrsim 110$\,\muhz, according to the stellar population study done by \cite{Miglio09}, with $\numax$ values provided by \cite{Huber11}. We then checked for available measurements of their g-mode period spacing $\Delta\Pi_1$, as measured by \cite{Mosser12}. This cross-match led to 21 stars observed by \kepler from Q0 till Q17.1, resulting in a frequency resolution of $\nu_\mathrm{bin}\simeq 0.008\,$\muhz\,\,for each power spectral density (PSD). The \kepler light curves were corrected according to \cite{Garcia14}.

In this work we focus on a low-mass low-luminosity RGB star, \kic, a particular type of RG that is part of a small population of stars (about 5\,\% of the entire population of RGs observed by \Kepler), having high signal-to-noise ratio (SNR) oscillations. Due to the clear oscillation pattern and the large number of modes, this star is well suited for testing stellar evolution theory and pulsation codes, as well as to constrain stellar formation rates in our galaxy. 


\section{Peak Bagging model and results}
\label{sec:2}
\begin{figure}
   \centering
\resizebox{0.50\columnwidth}{!}{%
   \includegraphics{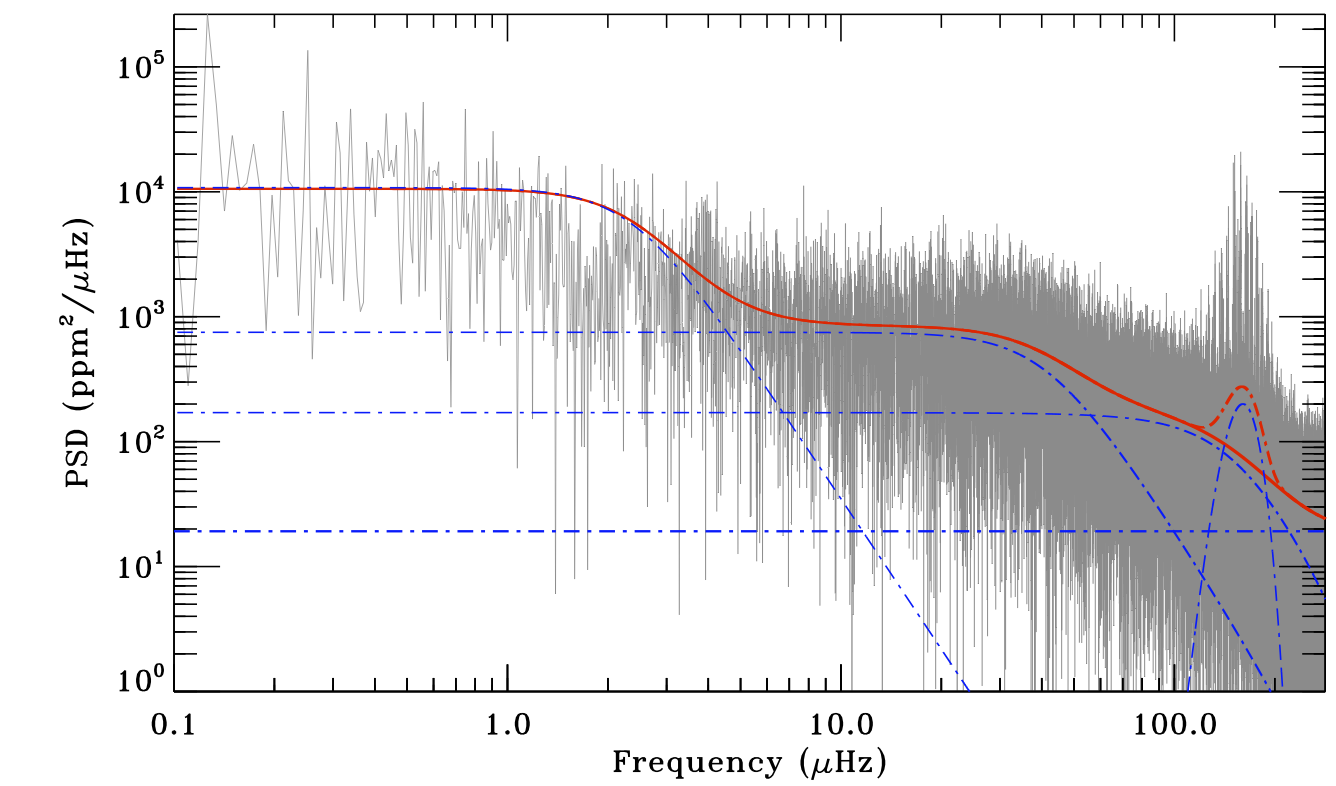}}
      \caption{Resulting background fit of the star KIC~12008916, as derived by \diamonds. The original PSD is shown in gray, while a smoothed version with boxcar width set to $\Delta\nu/5$, with $\Delta\nu$ taken from \cite{Huber11}, is shown as a black line to guide the eye. The red thick line represents the background model without the Gaussian envelope, while the red dotted line accounts for the additional Gaussian component. The individual components of the background model are shown by blue dot-dashed lines.}
    \label{fig:1}
\end{figure}

The first step for the peak bagging analysis of the stars is to estimate the background signal in their PSD. We follow the probabilistic study done on a large sample of stars by \cite{Kallinger14}, who found that two separate granulation components with a characteristic frequency close to the that of the region containing the oscillations, together with a third component related to long-term variations, best represent the background signal of the RGs selected. Thus, the background model that we adopt can be expressed as the sum of three super-Lorentzian profiles, a flat noise component, and a Gaussian envelope to model the power excess of the oscillations. 
The resulting background fit for KIC~12008916 is shown in Fig.~\ref{fig:1}.

The subsequent step in analyzing the stellar PSDs is to adopt an adequate fitting model for the oscillation pattern contained in the region of the power excess. In this work we adopt a method similar to that used by \cite{Corsaro14} (see Sect.~6.3). We restrict our frequency range of analysis to the region containing the oscillations, which we identify as $\numax \pm 3.5 \sigma_\mathrm{env}$, with $\sigma_\mathrm{env}$ the standard deviation from the Gaussian envelope used to fit the background. We thus take into account some substantial differences characterizing the oscillation pattern in RGs, as explained in the following. 

In the oscillation pattern of all the RGs, one can observe forests of modes between consecutive quadrupole and radial modes, which are known as dipole mixed modes, namely oscillations with a mixed character between that of pure $p$ and $g$ modes. Since $g$ modes have the highest mode inertia because they propagate in high-density regions, their lifetime $\tau$ is significantly longer than that of pure pressure modes (e.g. see \cite{CD04}). As a consequence, even with an observing time $T_\mathrm{obs} > 4$ years now made available from \Kepler, mixed modes with a g-dominated character appear still unresolved, while others having a more $p$ mode-like character are partially or sometimes even fully resolved. 

In this work we distinguish the resolved or partially resolved peaks, for which $T_\mathrm{obs} \gtrsim \tau$, from those that are  unresolved, for which $T_\mathrm{obs} \ll \tau$. In the former case, the oscillation peak profile that we adopt is that of a Lorentzian and it is given by
\begin{equation}
\mathcal{P}_{\mathrm{res},0} \left( \nu \right) = \frac{A_0^2 / \left( \pi \Gamma_0 \right)}{1 + 4 \left( \frac{\nu - \nu_{0}}{\Gamma_0} \right)^2} \, ,
\label{eq:resolved_profile}
\end{equation}
where $A_0$, $\Gamma_0$, $\nu_0$ are the amplitude, the linewidth and the frequency, respectively, and represent the three free parameters to be estimated during the fitting process.
According to the Fourier analysis, an oscillation peak that is not resolved has a profile represented as \cite{CD04}:
\begin{equation}
\mathcal{P}_{\mathrm{unres},0} \left(\nu \right) = H_0 \, \mbox{sinc}^2 \left[ \frac{\pi \left(\nu - \nu_0 \right)}{\nu_\mathrm{bin}} \right] \, ,
\label{eq:unresolved_profile}
\end{equation}
where $H_0$ and $\nu_0$ are the height in PSD units and the central frequency in \muhz\,\,of the oscillation peak, respectively, and must be estimated during the fitting process, while $\nu_\mathrm{bin}$ is the frequency resolution introduced in Sect.~\ref{sec:1} and it is a fixed value. In this case fitting the height is preferred since it is an observable and we have no information about the linewidth of the peak, while fitting a Lorentzian profile would represent an inadequate model as it involves a spurious fitting parameter represented by the linewidth. Then, the final peak bagging model can be represented as a mixture of Lorentzian and sinc$^2$ functions.

\begin{figure}
\centering
\resizebox{0.8\columnwidth}{!}{%
  \includegraphics{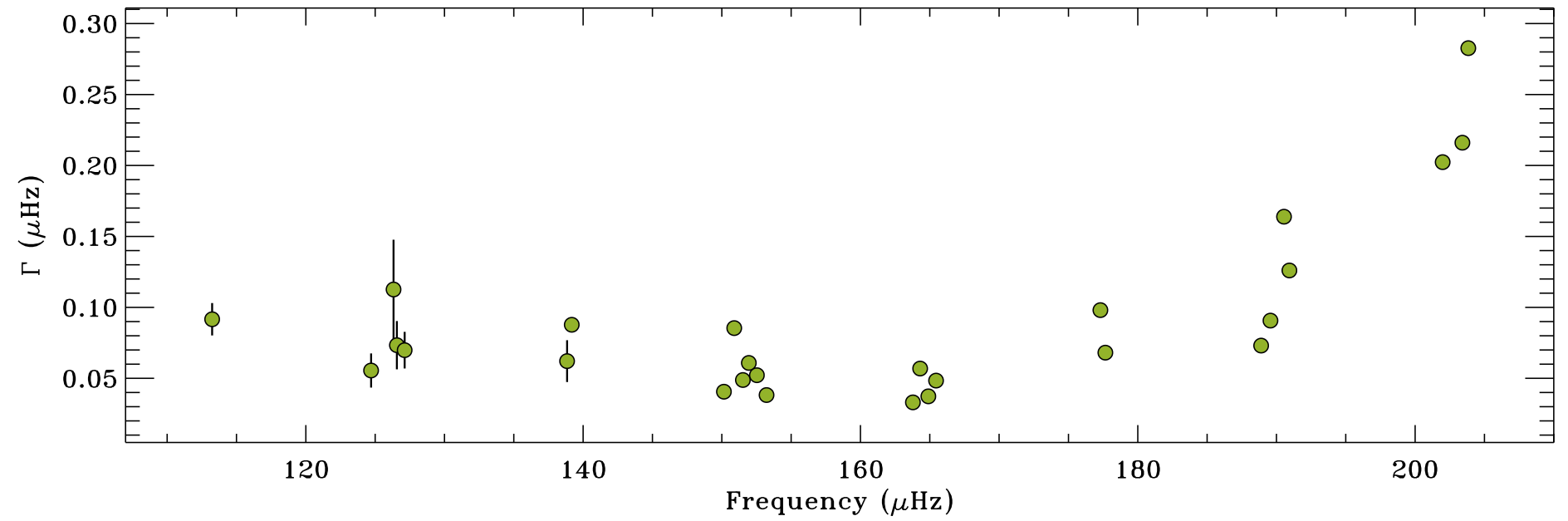}}
\caption{Linewidths of resolved dipole mixed modes for \kic\,\,as a function of the frequency in the PSD.}
\label{fig:2}       
\end{figure}

\begin{figure}
\centering
\resizebox{0.6\columnwidth}{!}{%
\includegraphics{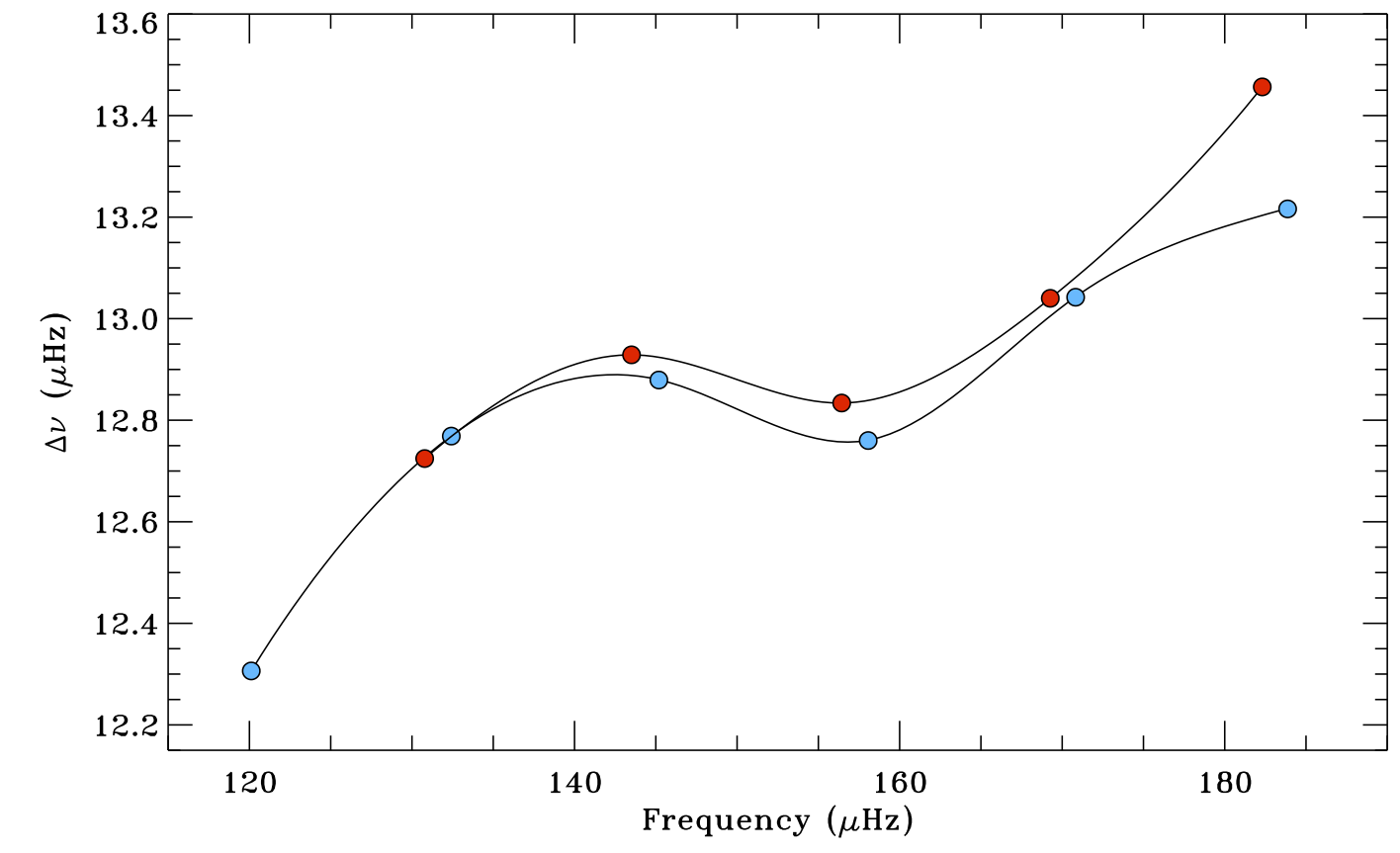}}
\caption{Signature of the acoustic glitches for \kic\,\,in the large frequency separation $\Delta\nu$ for the quadrupole and radial modes.}
\label{fig:3}       
\end{figure}

Some of the preliminary results for the peak bagging analysis of \kic\,\,are shown in Fig.~\ref{fig:2}, where one can observe a depression of the linewidths of dipole mixed modes in correspondence to the position of the frequency of maximum power $\numax$, and in Fig.~\ref{fig:3}, with the evidence of acoustic glitches for the quadrupole and radial modes.

\subsection{Peak significance}
\label{sec:3}
Testing the significance of an oscillation peak in the PSD of a star is a crucial aspect that has to be considered for providing a reliable set of modes that can be used for the modeling of the oscillations, hence for investigating the stellar structure and the evolution of the star. For this purpose, following \cite{Corsaro14} (Sect.~6.5) we consider a detection probability for a single peak, $p_\mathrm{B}$, defined as
\begin{equation}
p_\mathrm{B} \equiv \frac{\evid_\mathrm{B}}{\evid_\mathrm{A} + \evid_\mathrm{B}}
\label{eq:detection_probability}
\end{equation}
where $\evid_\mathrm{A}$ and $\evid_\mathrm{B}$ are the Bayesian evidences for the models excluding and including, respectively, the oscillation peak to be tested, and are computed with \diamonds. The detection limit is usually set to $p_\mathrm{B} \gtrsim 0.99$ as derived from a test involving a set of 1000 simulations of peak detection. In other words, we fit a chunk of PSD by considering on one side a model of the peak with the background and on the other side a model of the background alone. The best model selected with the Bayesian model comparison encompassed in Eq.~(\ref{eq:detection_probability}) is then telling us whether the peak is detected or not. 

The method presented in this work proves to be very efficient for this type of RG and will therefore be of great usefulness for the analysis of thousands of stars observed by \kepler.

%
%

\end{document}